\documentclass[journal]{IEEEtran}

\usepackage{amsmath} 
\usepackage{mathrsfs} 
\usepackage{multirow}
\usepackage{subcaption}
\usepackage{graphicx}

\begin{document}

\title{Studies on small charge packet transport in high-resistivity fully-depleted CCDs}
%
%

\author{Miguel~Sofo~Haro,
        Guillermo~Fernandez~Moroni,
        Javier Tiffenberg
\thanks{M. Sofo Haro is with: Centro Atómico Bariloche and Instituto Balseiro, Comisión Nacional de Energía Atómica (CNEA), Universidad Nacional de Cuyo (UNCUYO), Rio Negro, Argentina.}
\thanks{G. Fernandez Moroni is with: Instituto de Inv. en Ing. Eléctrica, Dept. de Ingeniería Eléctrica y de Computadoras, Dept. de Matemática, Universidad Nacional del Sur (UNS)-CONICET, Bahía Blanca, Argentina }
\thanks{The three authors are with Fermi National Accelerator Laboratory, United States of America, PO Box 500, Batavia IL, 60510}
\thanks{Manuscript received May XX, 2019; revised May XX, 2019.}}

%
%

\markboth{Journal of \LaTeX\ Class Files,~Vol.~14, No.~8, August~2015}%
{Shell \MakeLowercase{\textit{et al.}}: Bare Demo of IEEEtran.cls for IEEE Journals}

\maketitle

\begin{abstract}

In this work, we will present a physical model and measurements of the transport of small charge packets in the bulk of thick high resistivity CCD before being collected by the pixel potential wells. A new technique to measure the lateral spread of the charge as a function of the ionization depth in the bulk is presented. Results from measurements on CCD currently in use for several scientific instruments are shown and validated with a new mathematical algorithm to extend the current modeling based only on the diffusion of the charge in silicon. 

\end{abstract}


\IEEEpeerreviewmaketitle

\section{Introduction}

Although invented as memory devices \cite{Boyle_2010,Smith_2010}, Charge Coupled Devices (CCDs) have found a niche as imaging detectors due to their ability to obtain high resolution digital images of objects placed in its line of sight. In particular, scientific CCDs have been extensively used in ground and space-based astronomy and X-ray imaging \cite{Janesick_2001}. CCDs have high quantum efficiency, low read-out noise, good spatial resolution, and low dark current. Furthermore CCDs can be made thick and fully-depleted with high resistivity silicon to increase its detection mass, enabling their use as particle detectors \cite{holland2003fully}. 

\begin{figure}[t]
	\centering
	\includegraphics[width=\linewidth]{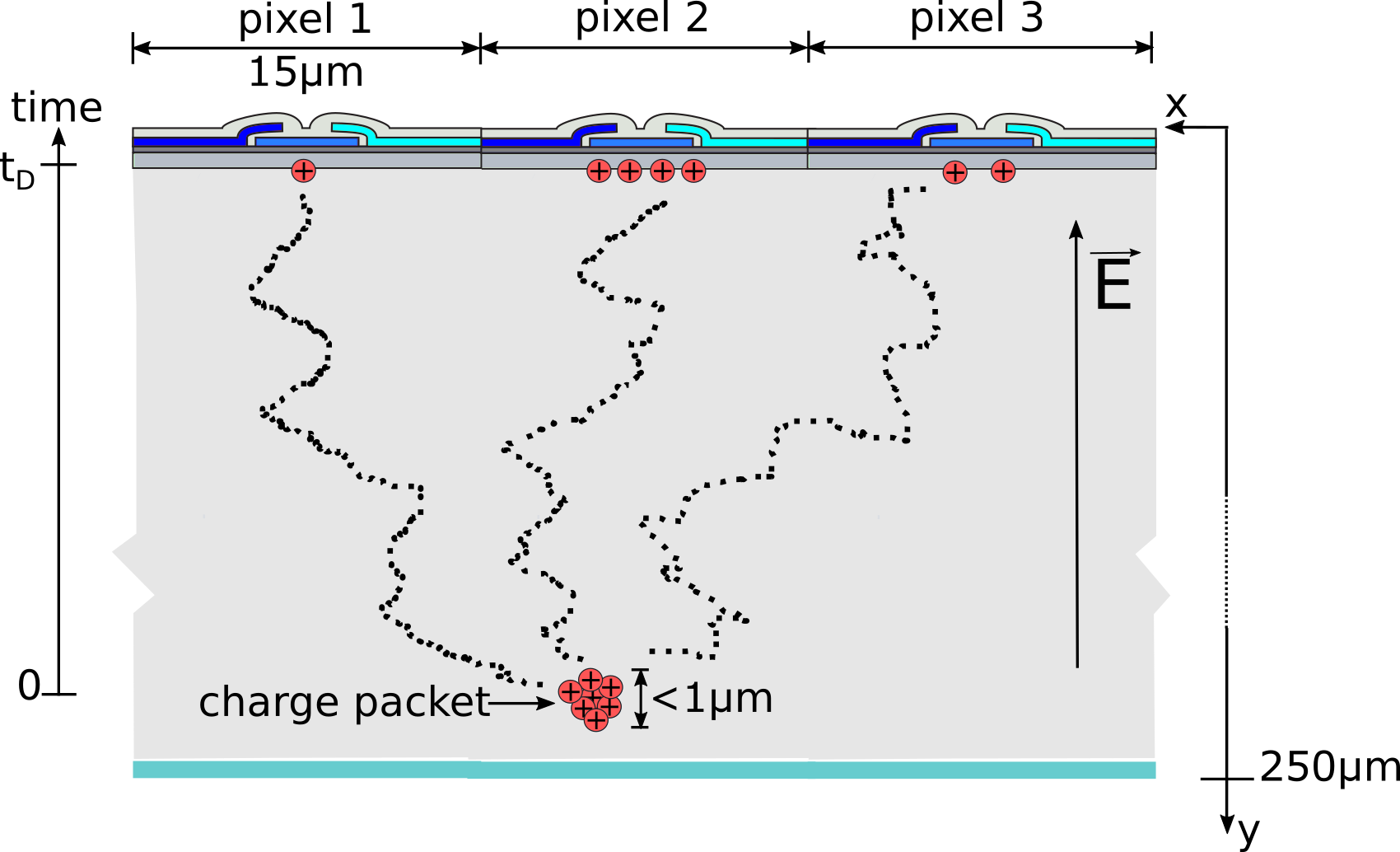}
	\caption{A small charge packet, with size less than $\rm 1\,\mu m$, is generated in the CCD bulk. Each charge carrier is drifted by the CCD electric field $\vec{E}$ to the collection well of the pixels. Due to diffusion and Coulomb repulsion effects, the charge carriers are spreaded over few pixels producing a pointlike event in the output image. In dashed lines is schematized the brownian trajectory followed by each one of the charge carriers. } 
	\label{fig:dibujito}
\end{figure}%

In particular, the low read-out noise of CCDs, around 2\,$\rm e^-$ makes possible set up a low energy threshold (RMS) of 5.5\,eV. Recently, with the development of thick fully-depleted skipper CCD, has been possible achieve a extremely low readout noise of 0.068$\rm e^-$ \cite{PhysRevLett.119.131802,moroni2012sub}. There are two novel experiments that takes advantage of this technology to detect the interaction between particles and the nucleus of the silicon atoms. The first is for direct dark matter detection called DAMIC \cite{Barreto_2012}, and the second one, is for neutrino detection called CONNIE  \cite{FernandezMoroni_2015}. In both cases, the particle scatters with one of the silicon nucleus transferring part of its kinetic energy to the crystal. The recoiling nucleus then produces an ionization cascade. Several electron-holes pairs are generated in the process in a volume much smaller than the pixel volume. A schema of the charge transport process that takes places in the CCD bulk can be seen in Figure \ref{fig:dibujito}. The ability to detect this signal above the noise floor (2\,$\rm e^-$) depends on how fast the carriers can spread out to neighbour pixels before being collected by the pixel well. The final size of the cloud is a monotonously increasing function of the depth of the charge generation point. Understanding and calibrating this relationship is a critical parameter to correctly address the detection efficiency on these experiments especially for small charge packet (below a few dozen of carriers) where the sensibility is higher. 

Due to the novel use of this devices for detecting small charge packet signals, there is no available technique to measure and model small charge packets transport in CCDs. Visible light application treats the charge transport only considering the carrier diffusion by the crystal thermal energy. This treatment is accurate since photons arrive individually to the sensor and produce one electron-hole pair each. In this scenario interactions between carriers (like repulsion) are negligible. In the other hand CCD users for X-ray detection has pushed some studies on the average behaviour of large charge packets, but only for certain photon energies and for thin CCD. In this article we present a novel technique to measure the charge packet size as a function of the depth for different number of carriers. We provide a mathematical model that we use to calibrate this relationship with measurements on a thick CCD.

The charge carriers are them spreaded by two process, and therefore collected in one or few pixels. One process, which is well understood in the field, is the diffusion process that occurs while the holes (or electrons, depending on the CCD type) are drifted towards the collection wells of the CCD pixels, and it is related with the depth of charge generation point in the bulk. As we will show in this work, the Coulomb repulsion among the charge carriers is another process with not negligible effect. In this work we will referred as \textbf{pointlike events} to any ionization in the CCD that takes place in a volume much smaller than the volume of a pixel. 


Section \ref{sec:ccds} is a description of the CCDs used in this work and in the former experiments. In section \ref{sec:imagePE} we will present the physical process that interplay in the transport of pointlike events and the final signature expected in the output image of the CCD. In section \ref{sec:mldlh} a maximum-likelihood estimator to measure the charge trasnport information from the events in the output images is presented. Section \ref{sec:simulation} details a simulation algorithm to model the charge movement. The new technique to measure charge packet spread as a function of bulk depth is explained in section \ref{sec:propTec}. In section \ref{sec:xrays} the technique is applied to an experiment with collimated X-rays. Then the experimental results are validated using the simulation algorithm, and the results are extrapolated to other charge packet sizes. Finally, we presents our conclusions in section \ref{sec:conclu}.

\section{High resistivity fully-depleted CCDs} \label{sec:ccds}

\begin{figure}[t]
	\resizebox{1\hsize}{!}{\includegraphics*{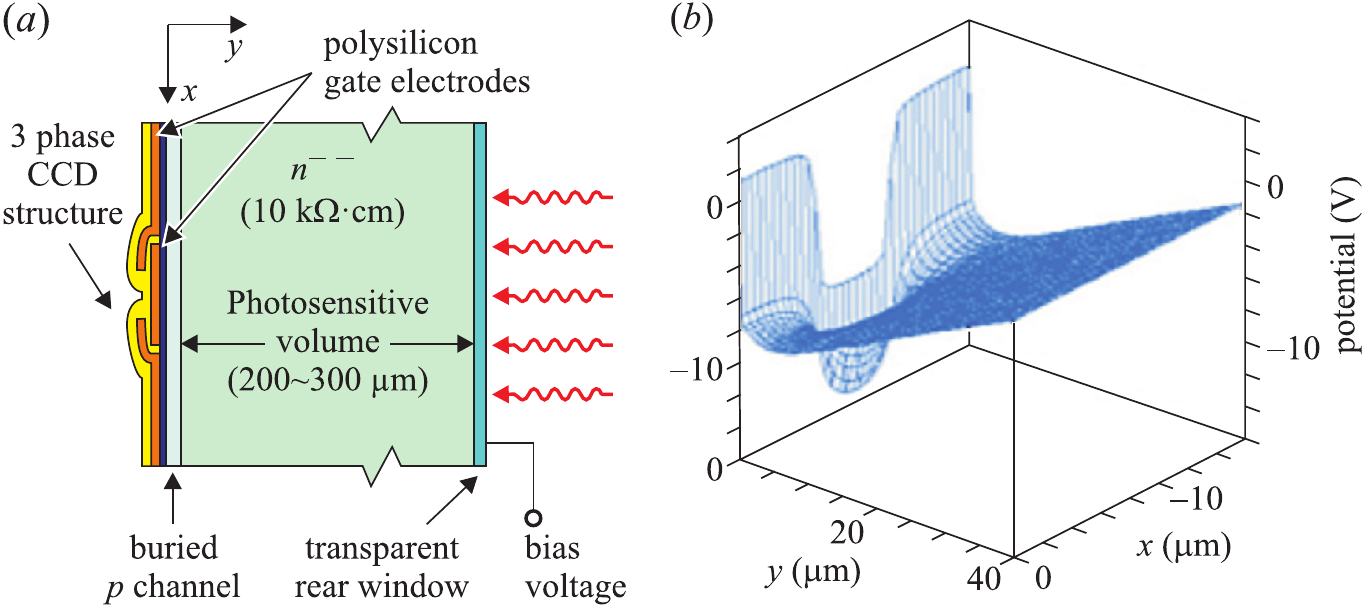}}
	\caption{Pixel cross section of a 250 $\mu$m thick CCD developed on Lawrence Berkeley Laboratory. Also, the electrostatic potential generated by the three phases under the gates is shown as function of depth ($Y$ axis) and one of the lateral directions ($X$ axis). Image extracted from \cite{holland2003fully}.} 
	\label{fig:CCD layout}
\end{figure}

The scientific CCDs used in this work has been developed by Lawrence Berkeley Laboratory and extensively characterized at Fermilab for the DECam project \cite{holland2003fully} \cite{Estrada_2006}. Their pixel structure is shown in figure  (\ref{fig:CCD layout}). Several million of CCD pixels of $\rm 15\,\times15\,\mu m^2$ are fabricated on high resistivity silicon to maximize the depleted silicon volume and therefore increase the near-IR photon response. The CCDs are three-phase, p-channel, back illuminated with a special coating in the backside to increase photon absorption and to reduce back side dark current generation. However, for dark matter and neutrino detection the optical features are not relevant. 


The readout noise is the main source of error for the actual value of each pixel in the output image \cite{haro2016measurement}. The noise is added by the output amplifier when the pixel charge packet is read, and affects every pixel \cite{haro2016taking}. The readout error produced by this noise is a normally distributed random variable with zero mean and standard deviation ($\sigma_R$) that depends on the time spent to read each pixel (readout time) with a minimum of $\sim$2\,$\rm e^-$ \cite{Cancelo_2012}. There is almost no correlation among errors of different pixels, and because of their normal distribution, each error sample can be considered independent of the others \cite{moroniRPIC2015}.

\section{Image produced by small charge packets} \label{sec:imagePE}

\begin{figure}[t]
	\centering
	\includegraphics[width=0.8\linewidth]{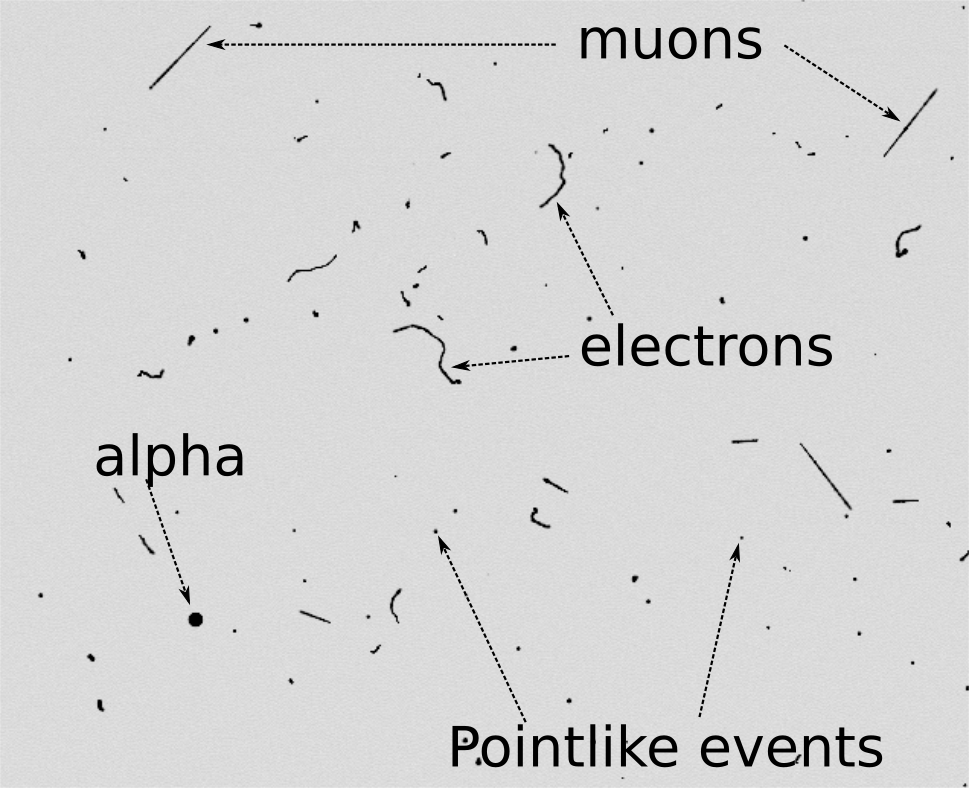}
	\caption{Compendium of particle events at sea level using a 250\,$\rm \mu m$ thick fully-depleted CCD. For details see text.} 
	\label{fig:compendium of background events}
\end{figure}

This section addresses the physical processes that participate in the transport of free carriers in small charge packets. These events are generally produced by particles that interacts with silicon atoms and transfer some or all of their energy to the crystal net. Figure (\ref{fig:compendium of background events}) is an output image from a blind measurements with a CCD at sea level, where it is possible to observe a compendium particles interactions. The black signatures are ionization tracks from different energetic particles as muons, electrons and alpha produce events that occupy many pixels in the final image. The muon track appears as a straight line crossing the entire silicon volume; energetic electrons from electromagnetic radiation produce wiggling tracks like worms; big and bright dots are produced by alpha particles because the plasma effect that they generate in the silicon \cite{Estrada_2011}. In the other hand you also have low energy depositions that we call pointlike events where their final shape in the output image is dominated by the charge transport processes in the silicon and not by the particle interaction. 

\subsection{Initial cloud size} \label{sec:pesize}

\begin{figure}[t]
	\resizebox{1\hsize}{!}
	{\includegraphics*{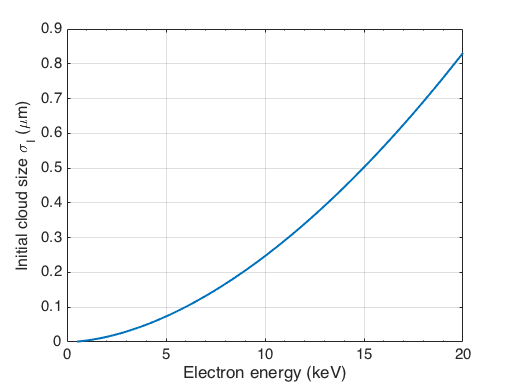}}
	\caption{Initial cloud size $\sigma_I$ of the pointlike event for different energies of the primary electron released by an X-ray.} 
	\label{fig:rangoElectrones}
\end{figure}

The most probable particle producing a pointlike event with energy in the keV range is the photon. They also provide an easy way to generate small charge packets in the laboratory. In particular in our experimental section we use X-rays with energies below 20\,keV. The most probable interaction mechanism of these photons with silicon is the photoelectric effect \cite{knoll2010radiation} where all its energy $E_X$ is absorbed by only one electron. This electron is released from the atom with an energy $E_{e^-}=E_X-E_B$, where $E_B$ is the binding energy of the electron. The free photo-electron can then produce extra ionization. More details about the secondary ionization processes can be found in \cite{Janesick_2001,moody2017ccd}. The average number of free carriers is given by \cite{Janesick_2001}

\begin{equation}
	N_{e^-}=E_{e^-}/w_{e^-}.
	\label{eq:number_electrons}
\end{equation}

where $w_{e^-}$ is the average energy required to generate one electron-hole pair in silicon, it is equal to 3.77\,${\rm eV}$ at the CCD temperature operation of 140\,kelvin \cite{groom2004temperature}. The initial position of the carriers can be modeled following a 3D Gaussian distribution with a deviation $\sigma_I$ given by $0.257R_{e^-}$ \cite{yousef2011energy}, where $R_{e^-}$ is the range of the primary electron given by,  

\begin{equation}
	R_{e^-} (\mu m)=0.0171E_{e^-}^{1.75}
	\label{eq:electronRange}
\end{equation}

where $E_{e^-}$ is in keV. Figure \ref{fig:rangoElectrones} shows the $\sigma_I$ for different $E_{e^-}$. For example, a photo-electron of 10\,keV will produce a ionization cloud with standard deviation size of less than 0.3\,${\rm \mu m}$ \cite{williams2001x,ashley1976calculations,moody2017ccd}. Therefore, the x-rays can be used as test particles to characterize the image produced by pointlike events. 

\subsection{Charge movement by the CCD electric field} \label{seq:electric}

Once the electron-hole pairs are created, they are drifted by the electric field in the depleted region of the CCD and for the p-channel CCD (as used in this work) the holes are pushed towards the potential wells under the gates. The simulation of the electric potential in figure (\ref{fig:CCD layout}) shows that the well extends only a few $\rm \mu$m in depth. Beyond the first 15\,$\rm \mu$m, there is no appreciable lateral barrier along the $x$ axis (and along the third dimension $z$ which is not shown in the figure). The potential only varies with depth ($y$ axis). Therefore the electric field in the bulk can be modeled as a linear function of $y$. In particular, as the dopant concentration $N_D$ is uniform in this region, the electric field is given by
\begin{equation}
\label{eq:electric field}
E(y)=a_1y-a_2,
\end{equation}
with
\begin{equation}
    a_1=\frac{qN_D}{\epsilon_{si}}
    \label{eq:a_1}
\end{equation}
\begin{equation}
    a_2=\frac{qN_D}{\epsilon_{si}}y_J+E_J
    \label{eq:a_2}
\end{equation}

where $q$ is the electron charge, $\epsilon_{si}$ is the dielectric constant of the silicon, $y_J$ the thickness of the p-channel, and $E_J$ is the field in the p-n junction. The derivation of this model can be found in \cite{holland2003fully}.

The average drift velocity ($v$) of the free charge is proportional to the electric field: $v(y)=\mu E(y)$, where $\mu$ is the hole mobility. The hole mobility is a function of the electric field and can be empirically modeled from \cite{canali1975electron} as

\begin{equation}
\mu(E,T)=\frac{1.31 \times 10^8\,T^{-2.2}}{\left[1+\left(\frac{E}{1.24\,T^{1.68}}\right)^{0.46\,T^{0.17}}\right]^{1/(0.46\,T^{0.17})}}
\label{eq:movilidad}
\end{equation}

\begin{figure}[t]
	\resizebox{1\hsize}{!}
	{\includegraphics*{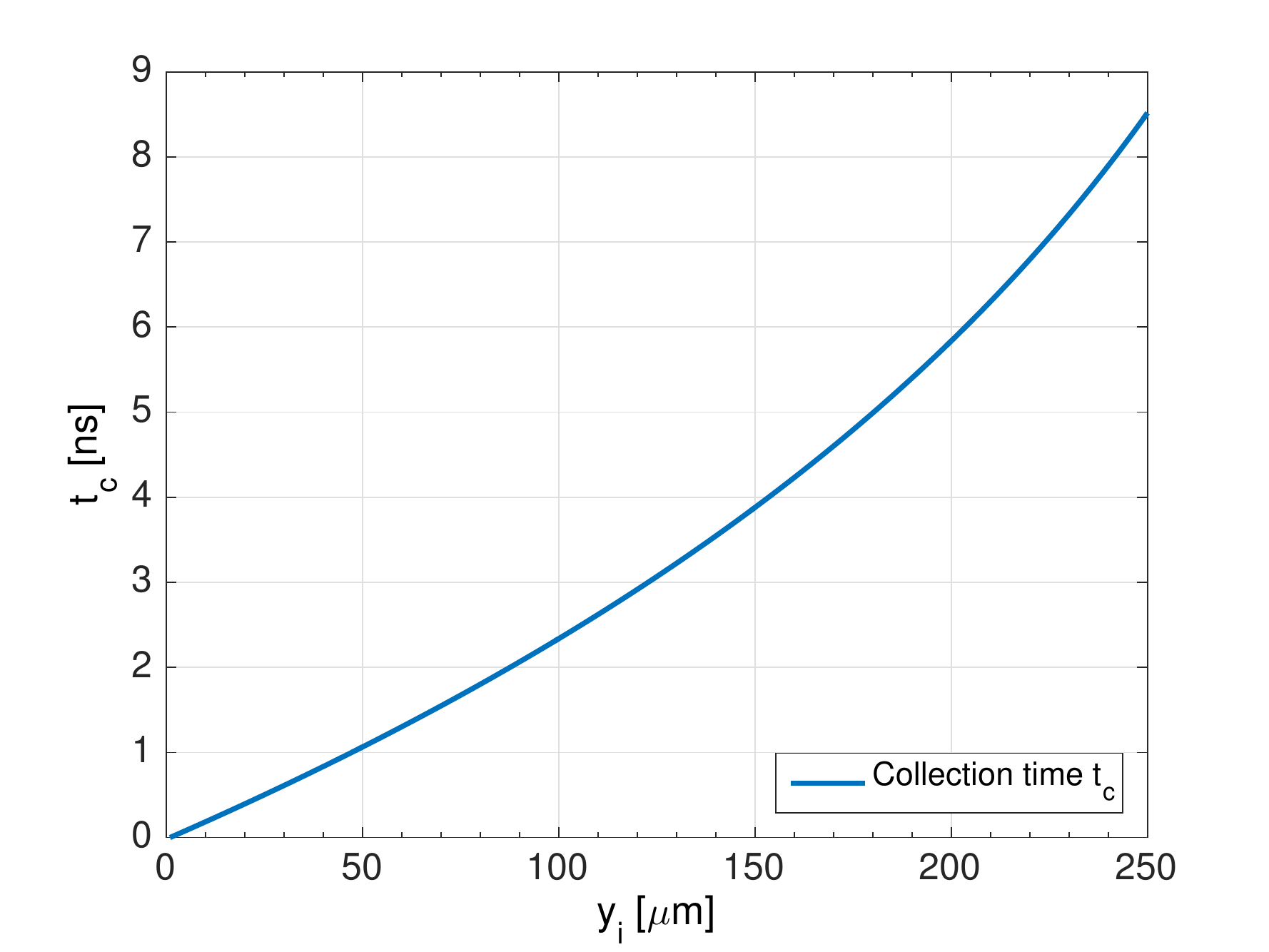}}
	\caption{Collection time of an event produced at depth $y_i$. It was obtained numerically solving equation (\ref{eq:diffTreco}) with the mobility of equation \ref{eq:movilidad}.} 
	\label{fig:coltime}
\end{figure}

The collection time $t_c$ needed to move a hole to the potential well, can be obtained solving the equation (\ref{eq:diffTreco})
\begin{equation}
\int_{y_i}^{y_J} \frac{1}{v(y)}dy=\int_{t=0}^{t_c}dt
\label{eq:diffTreco}
\end{equation}
where $y_i$ is the depth where the charge was generated and $y_J$ the potential well depth. Fig. (\ref{fig:coltime}) shows the collection time as a function of the generation depth point. 

\subsection{Charge diffusion} \label{sec:diffusion}

Free carriers generated in the bulk of the CCD do not necessarily end up being collected by the corresponding pixel at that position. Some of the free charge moves to neighbor pixels by diffusion. The diffusion movement is produced by the elastic scattering of the charge with atoms of the net, which have the thermal kinetic energy of the media. The direction and velocity after each collision are random since the collision is dominated by the kinematics of the net (due to the difference in momentum between the atom and electron). The average free path among collisions is very small compared to the pixel volume and therefore the charge movement can be characterized as a Brownian motion process. The carriers are free to diffuse until they are collected by the potential well of the pixels.

If the position of the free charge in the array after a time $t_D$ is described by the random variables $X_D$, $Y_D$ and $Z_D$, modeled as a Brownian Motion process \cite{Evans_2012}, their joint distribution is normal, $N(\vec{\mu}_g,\vec{\sigma}_D)$, where $\vec{\mu}_g=(x_g,y_g,z_g)$ are the coordinates of the electron-hole pair ionization point, and $\vec{\sigma}_D=\mbox{diag}(\sigma_D,\sigma_D,\sigma_D)$ is the diffusion movement of the charge in both directions (which are assumed to be equal). The projection of the displacement in the $x$-$z$ directions are relevant for this work since this is the information available in the output images. From Einstein's equation of diffusion \cite{Einstein_1956}, the variance of the traveled path can be calculated as
\begin{equation}
\sigma_D^2=2D_ht_D
\label{eq:difu}
\end{equation}
where $D_h$ is the holes diffusion coefficient given by
\begin{equation}
D_h=\frac{k_B\mu_hT}{q}
\label{eq:diffusion}    
\end{equation}
where $k_B$ the Boltzmann constant and $q$ the electron charge. For a charge packet generated at depth $y_i$, the diffusion effect is obtained using the collection time from Figure \ref{fig:coltime}. 


%
%

\subsection{Charge Coulomb repulsion} \label{sec:teoria_repulsion}

Another factor that affects the event charge spread in fully-depleted CCDs is the Coulomb repulsion. Since typical applications with thick CCD are used for visible light or infrared light that produces one electron-hole pair per photon, this process has not been incorporated in the analysis for the lateral movement of the charge. However other similar silicon-based technologies have shown that its effect is not negligible  \cite{yousef2011energy,gatti1987dynamics,kimmel2010experimental,castoldi20123,mutoh2016device,rykov1994ssd}. 
 
 The created holes can be treated as a cloud of free carriers that repel each other in their way to the pixel well. For a cloud of N holes, the mutual electric field contribution on the hole $j$ located in a position $\vec{r}_j=(x_j,y_j,z_j)$ in the crystal will be given by

\begin{equation}
\vec{E}(\vec{r}_j) = \sum_{i=1}^N \vec{E}_i(\vec{r}_i) =
\frac{q}{4\pi\varepsilon_{Si}} \sum_{i=1}^N \frac{\vec{r}-\vec{r}_i}{|\vec{r}-\vec{r}_i|^3}
\label{eq:campoElectrico}
\end{equation}

where $\vec{r}_i$ is the position of the hole $i$. This process will have larger effect at the generation point where the carriers are closer together. 
The velocity for each carrier ($j$) is proportional to the electric field vector using the mobility of equation \ref{eq:movilidad} $$\vec{v}_j=\mu_j(E_{j},T)\vec{E}(\vec{r}_j)$$ where  $E_j =|\vec{E}(\vec{r}_j)|$.

\section{Numerical charge transport algorithm}
\label{sec:simulation}
The dynamics of the charge carriers is an interplay among diffusion, drift and Coulomb repulsion. It is described by a stochastic differential equation \cite{gatti1987dynamics}, that has not been solved in closed form \cite{rykov1994ssd}. In order to validate and extrapolate the experimental results, a Monte Carlo algorithm was developed based in previous work from Castoldi et al \cite{castoldi20123}. The final outcome of the algorithm is the statistical properties of the position of the charge when they are collected by the potential well of the pixels after the transport in the detector bulk. The algorithm can be described in three steps:

\begin{enumerate}
    \item A charge packet with a given number of carriers ($N_c$) is generated following and initial photon interaction in eq. \ref{eq:number_electrons}. The initial position (at time 0) of a carrier $j$ ($\vec{r}_{j,0}=(x_j,y_j,z_j)$) is obtained by a random generator following a normal distribution with mean as the desired mean position of the interaction $\vec{\mu}_g = (x_g, y_g, z_g)$ and standard deviation $\vec{\sigma_g} = \mbox{diag}(\sigma_g, \sigma_g, \sigma_g)$ with $\sigma_g=0.257R_{e^-}$ from section. \ref{eq:electronRange}: $\vec{r}_{j,0} \sim N(\vec{\mu},\vec{\sigma_g})$.
    
    \item The evolution of the carriers in the bulk are then evaluated using a time step simulation. On each time step ($s$), the position of the carriers ($\vec{r}_{j,s}=(x_{j,s}, y_{j,s}, z_{j,s})$) is updated by the diffusion, drift and repulsion processes. This is repeated until the position in the $y$ direction reaches the potential well edge ($y_w$). At that time the position of the carrier is recorded and its contribution to the repulsion to the other carriers is eliminated. 
    
     At each time step $s$, the electric field vector $\vec{E}_{j,s}$ ($j = 1\dots N_c$) for all the holes is calculated as the superposition of CCD electrostatic field (eq. \ref{eq:electric field}) and the contribution from the repulsion from eq. \ref{eq:campoElectrico}. Using $E_{j,s}=|\vec{E}_{j,s}|$, the mobility $\mu_{j,s}(E_{j,s},T)$ from eq. \ref{eq:movilidad} is calculated for each hole. The velocity due to the electric field is $\vec{v}_{j,s}=\mu_{j,s}\vec{E}_{j,s}$ and the change in position is calculated as $\vec{dr}^r_{j,s}=\vec{v}_{j,s} dt$. 
     
     In the same time step, a normal distributed random displacement due to diffusion with variance $\sqrt{2D_h\Delta t}$ is calculated for each hole: $\vec{dr}^d_{j,s}\sim N(\vec{0},\vec{\sigma_D})$ ($\sigma_D$ from section \ref{sec:diffusion}).
     
    The position is updated by $\vec{r}_{j,s+1} = \vec{r}_{j,s}+\vec{dr}^r_{j,s}+\vec{dr}^d_{j,s}$. If $y_{j,s+1}<y_w$ then the position of the hole is recorded and removed for the following iterations. The time is updated $t_{s+1} = t_s + dt$.
    
    \item The final position recorded for all the carriers is used to calculate the standard deviation in the $x$ and $z$ directions and other statistical properties of the charge packet transport. 
\end{enumerate}

The time step $dt$ should be small so the assumption that the electric field for each carrier is constant during the time step is a good approximation of the real scenario. 

\begin{figure}[t]
	\centering
	\begin{subfigure}[t]{\linewidth}	
		\centering	
		\subcaption[]{$\rm x_i=2.10$, $\rm z_i=2.30$, $\rm y_i=50\,\mu m$, $\sigma_s=0.2$\,pixel-size}	
		\includegraphics[width=\linewidth]{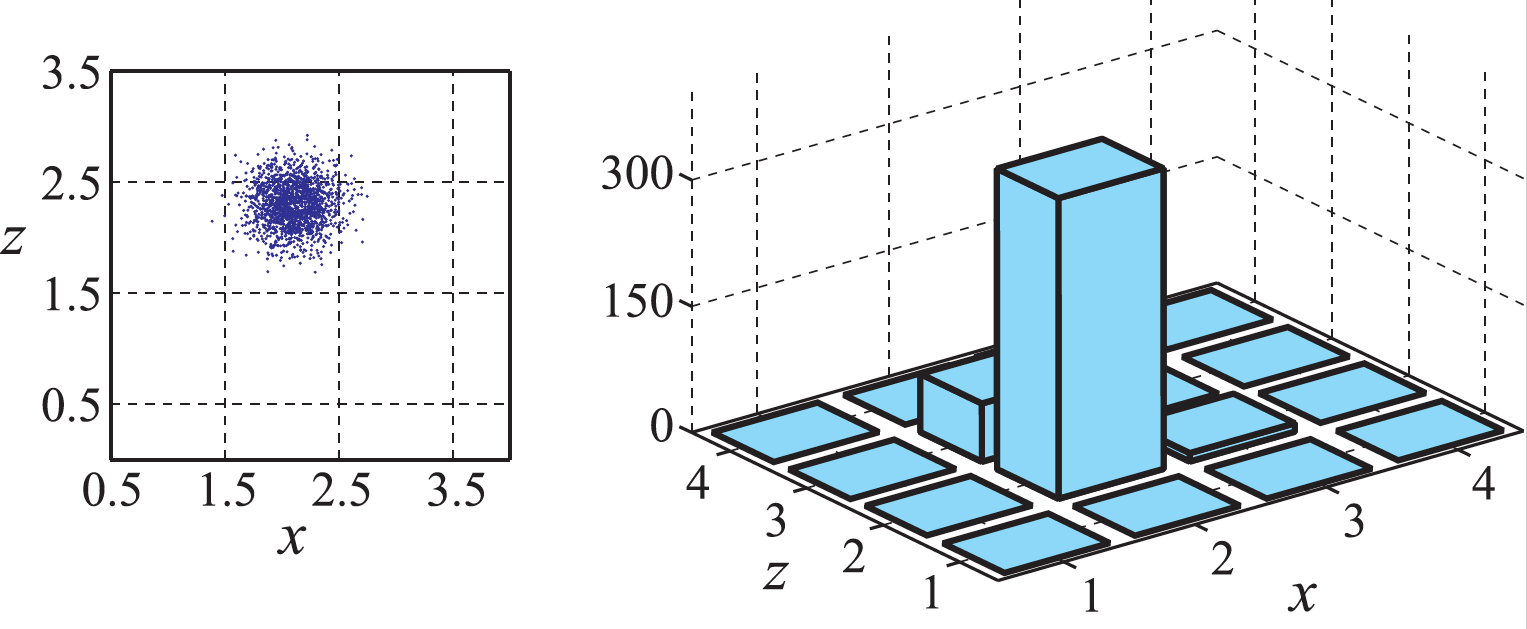}
		\label{fig:ejEventTop}
	\end{subfigure}
	\begin{subfigure}[t]{\linewidth}
		\centering
		\subcaption[]{$\rm x_i=2.25$, $z_i=1.60$, $\rm y_i=250\,\mu m$, $\sigma_s=0.5$\,pixel-size}	
		\includegraphics[width=\linewidth]{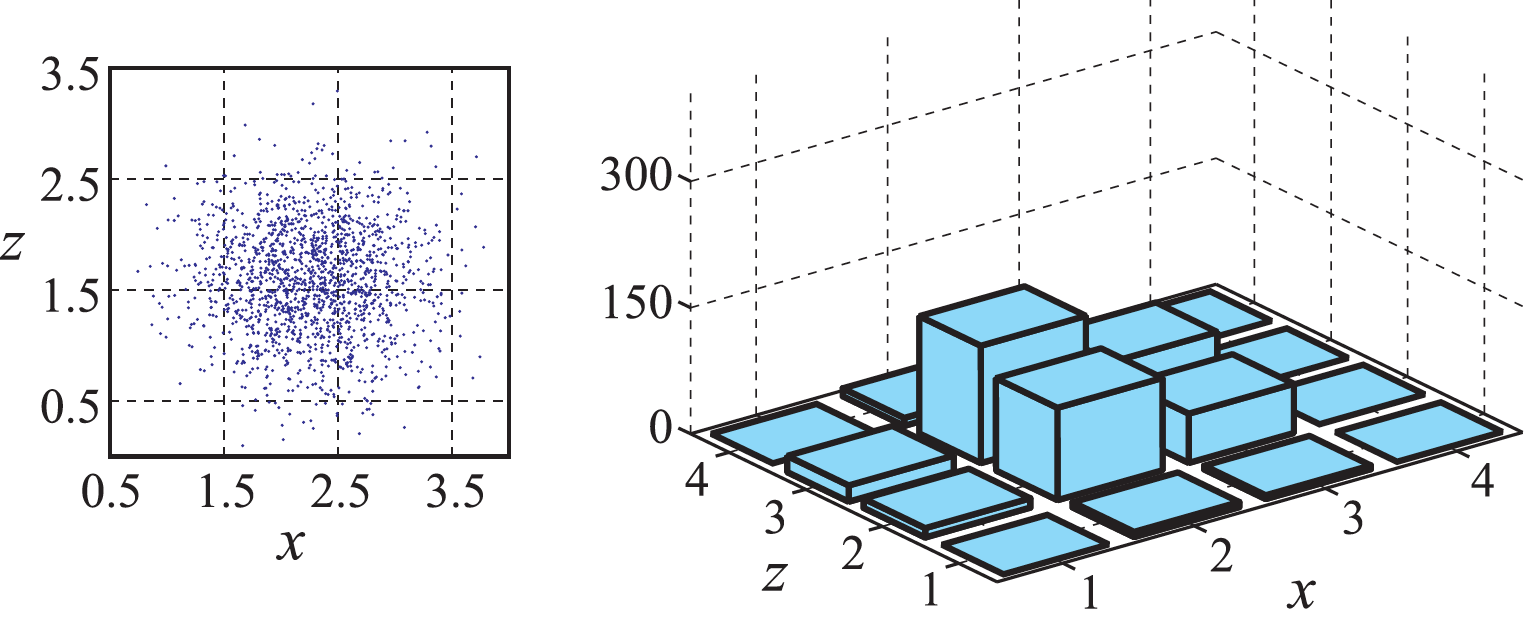}
		\label{fig:ejEventBot}
	\end{subfigure}
	\caption{Simulation of 500 holes charge packet at two different positions in the array and different depth. The generation point of the event of figure (\subref{fig:ejEventTop}) was at a depth close to the potential well of the pixels, and the event of figure (\subref{fig:ejEventBot}) at a deeper position.} 
	\label{fig:simulated events}
\end{figure}

After all carriers are simulated through the silicon it is possible to quantize their position to the pixels coordinates. The final pixel value of the simulation is the sum of all the carriers inside the boundaries of one pixel. A Gaussian noise can then be added to this value to simulate the readout noise of the system. Figure \ref{fig:simulated events} shows a simulation of two charge packets generated at different depth in the silicon.

\section{Measurement of the carriers spread from output images} \label{sec:mldlh}

After reading the CCD, the events from the output images are extracted. The pixels with charge above a given threshold that are sharing at least one edge are classified for the same event. This pixels are saved together with the first line of pixels surrounding those with value above the threshold. The information for all this pixels for one event is then used to estimate dispersion of the carriers before being collected by the pixel wells. Since the expected RMS dispersion is smaller than the pixel size, there is a quantization of the carrier position by the geometrical limits of the pixel in the array. The expected 2D density probability function of the pixels in the image by the collection of a small charge packets is (from \cite{moroniRPIC2015})
\begin{equation}
\label{eq: PDF evento}
\begin{array}{l}
f_{\textbf{P}}(\textbf{p};N_c,x_g,z_g,\sigma_s)=\medskip \\
\prod\limits_{i=1}^{N_p}\sum_{q_i=0}^{N_c}\frac{N_c!\lambda_i^{q_i}(1-\lambda_i)^{N_c-q_i}}{q_i!(N_c-q_i)!}\frac{e^{-\frac{(p_i-q_i)^2}{2\sigma_R^2}}}{\sigma_R\sqrt{2\pi}}
\end{array}
\end{equation}
where $\textbf{P}=(P_1,...,P_{N_p})^T$ is the vector of random variables of the pixels value of the event, $N_c$ is the total charge of the event, $\sigma_R$ is the standard deviation of the readout noise explained above, $N_p$ is the number of pixels in the events, and $(x_g,y_g,z_g)$ are the coordinates in the volume of the detector where the charge was originated, $\sigma_s$ is the standard deviation that measures the spread of the free carriers before being trapped by the potential wells, and $\lambda_i$ is the probability that one hole of the pointlike event ends in the pixel $i$ given by
\begin{equation}
\label{eq:pixel probability of charge}
\lambda_i=\! \int_{x_{i,o}}^{x_{i,f}} \! \int_{z_{i,o}}^{z_{i,f}}\frac{e^{-[(x-x_g)^2 + (z-z_g)^2]/(2\sigma_{s}^2)}}{2\pi\sigma_{s}^2}dz dx
\end{equation}
where $x_{i,0},x_{i,f}, z_{i,o},z_{i,f}$ are the positions of the borders of the $i$-th pixel of the array. 

\begin{figure}[t]
	\centering
	\includegraphics[width=0.8\linewidth]{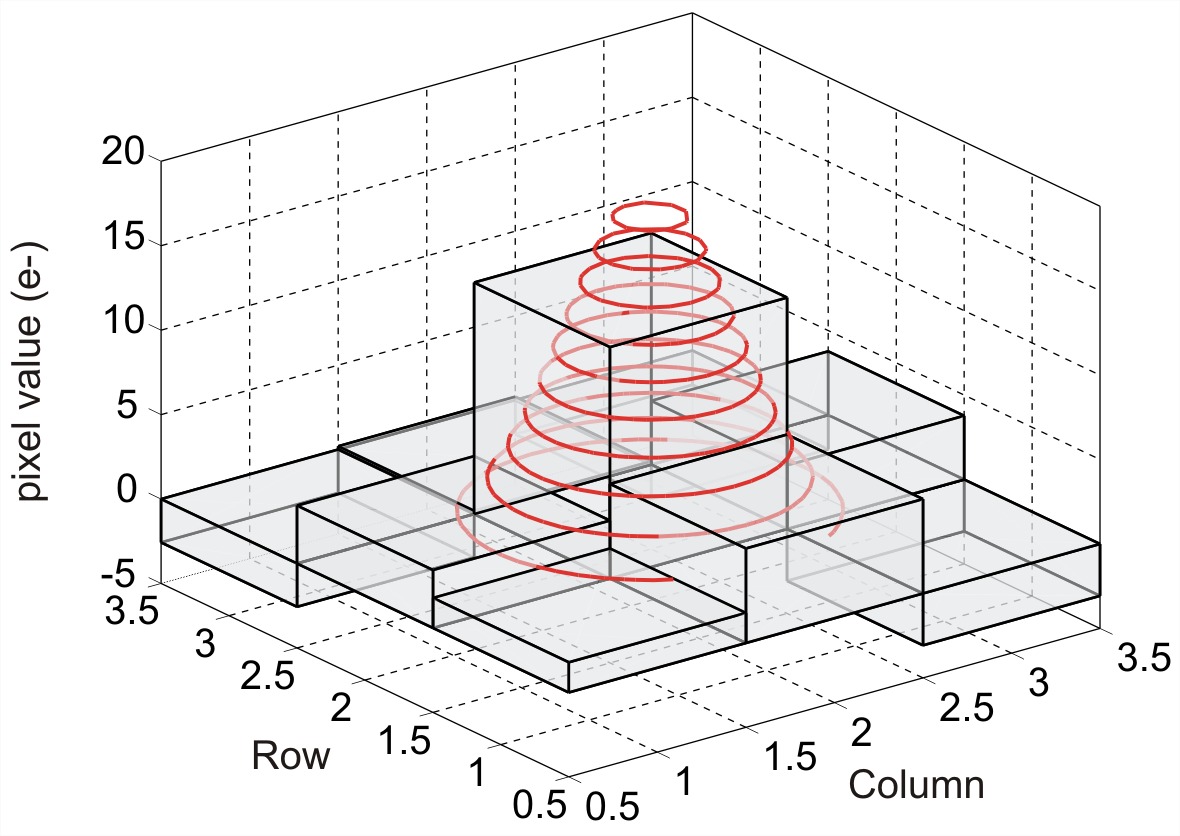}
	\caption{Example of a likelihood fit for a 20\,$\rm e^-$ event. In grey are the pointlike event pixels, and in red is the likelihood function fit with the events pixels.} 
	\label{fig:fitexample}
\end{figure}

Then, $f_{\textbf{P}}$ can be used to estimate the spread the original spread of the carriers ($\sigma_s$) using a likelihood estimator
\begin{equation}
L(N_c,x_g,z_g,\sigma_s)=\frac{max}{N_c,x_g,z_g,\sigma_s}f_{\textbf{P}}(\textbf{p};N_c,x_g,z_g,\sigma_s)
\label{eq:likelihood}
\end{equation}
For this work the maximization of the algorithm is implemented using the MINUIT tool with the SEEK option that applies a Monte Carlo search algorithm \cite{james1975minuit}. The events are detected from the image using a seed given by any pixel with a value higher than four times the standard deviation of the readout noise. Any adjacent pixel with value three times higher than the readout noise is also added to the event. In figure \ref{fig:fitexample} there is an example of a pointlike event of 20\,$\rm e^-$ with its corresponding likelihood fit. 

\section{Proposed technique to measure the $\sigma_s$-$y_g$ relationship} \label{sec:propTec}

This section provides the mean to measure the dependence of spread of the carriers in the silicon as a function of the generation depth of the small charge packet. 
The technique uses an interacting particle with a known probability of interaction as a function of the depth in the detector. The theoretical deposition is then compared to the measured profile of event sizes ($\sigma_s$) in the output images. In particular, this work uses X-rays with several keV of energy as the prove particle because they are easy to generate, its energy is very well known and completely absorbed by an electron through the photoelectric effect. 

Given a particle that generates a small charge packet at a depth $y$ following a known cumulative probability distribution $G^F_Y(y)$, the cumulative probability of the spread function of the event is
\begin{equation}
G^F_{\sigma_s}(\sigma_s)=G^F_Y(\sigma_s(y))
\label{eq:cumulative diffusion}
\end{equation}
where $\sigma_s(y)$ is the function of the spread of the carriers before being collected by the pixels as a function of the interaction depth. It is the function that we want to estimate with the method. The super-index $F$ in the distributions means that the distributions are calculated starting the depth in silicon from the front of the detector. Similar results can be derived if the distributions are calculated from the back. A realization of the distribution $G^F_{\sigma_s}(\sigma_s)$ can be measured from the events in the output images of the experiment as 
\begin{equation}
\hat{G}^F_{\sigma_s}(\sigma_s) = \frac{N^F_e(\sigma_s)}{N_TG^F_Y(y_w)}
\label{eq:cumulative}
\end{equation}
where $N^F_e(\sigma_s)$ is the number of events with spread from 0 to $\sigma_s$, $N_T$ is the total number of events detected and $G^F_Y(y_w)$ is the known theoretical cumulative deposition distribution evaluated at the CCD edge. From equation (\ref{eq:cumulative diffusion}), it is possible to calculate the value of $y$ for each spread as:
\begin{equation}
\label{eq: y from measurements}
y={G^F_Y}^{-1}(\hat{G}^F_{\sigma_s}(\sigma_s)).
\end{equation}

Any pair $(y,\sigma_{s})$ that solves eq. (\ref{eq: y from measurements}) is a point of the calibration of the $\sigma_s(y)$ function.

\subsection{$G_Y(y)$ for X-rays}  \label{sec:GYyXrays}

\begin{figure}[t]
	\centering
	\includegraphics[width=5cm]{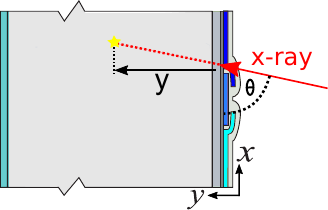}
	\caption{X-ray with an incident angle $\Theta$ and a interaction depth of $y$.} 
	\label{fig:geometría de rayos}
\end{figure}

For an X-ray of a given energy entering to the CCD with an angle $\theta$ to the surface of the detector (0$\le \theta \ge \Pi/2$), as shown in figure \ref{fig:geometría de rayos}, its probability to reach a depth of $y$ measured from the surface of the detector is
\begin{equation}
\label{eq: distribution from theta}
g_Y(y;\theta)=\frac{e^{-\frac{y}{\gamma\:\text{sin}(\theta)}}}{\gamma\: \text{sin}(\theta)}, \: y\geq 0.
\end{equation}

where $\gamma$ is the attenuation length of the X-ray in silicon, which is energy dependent. In the same way, the probability to interact before a depth $y$ is
\begin{equation}
\label{eq: cumulative distribution from density}
G_Y(y;\theta)= 1-e^{-\frac{y}{\gamma\:\text{sin}(\theta)}}, \: y\geq 0.
\end{equation}

In a more general approach, the X-rays can reach the detector at a different angles with a given distribution $g_{\Theta}(\theta)$. In this case, the joint probability distribution of $Y$ and $\Theta$ can be calculated as
\begin{equation}
g_Y(y,\theta)=\frac{e^{-\frac{y}{\gamma\:\text{sin}(\theta)}}}{\gamma\: \text{sin}(\theta)}g_{\Theta}(\theta), \: y\geq 0,\: 0\leq \theta \leq\pi/2,
\end{equation}

Then, the marginal distribution for $Y$ and its cumulative is determined by
\begin{equation}
g_Y(y)=\int\limits_{0}^{\pi/2}\frac{e^{-\frac{y}{\gamma\:\text{sin}(\theta)}}}{\gamma\: \text{sin}(\theta)}g_{\Theta}(\theta)d\theta, \: y\geq 0.
\end{equation}
\begin{equation}
G_Y(y)=\int\limits_{0}^{y}g_Y(u)du,
\end{equation}
\section{Experimental results}

In this section we present an experiment to apply the measurement technique of section \ref{sec:propTec}. The CCDs used in the experiment was the one presented in section \ref{sec:ccds}. 

\begin{figure}	
	\centering		
	\includegraphics[width=0.8\linewidth]{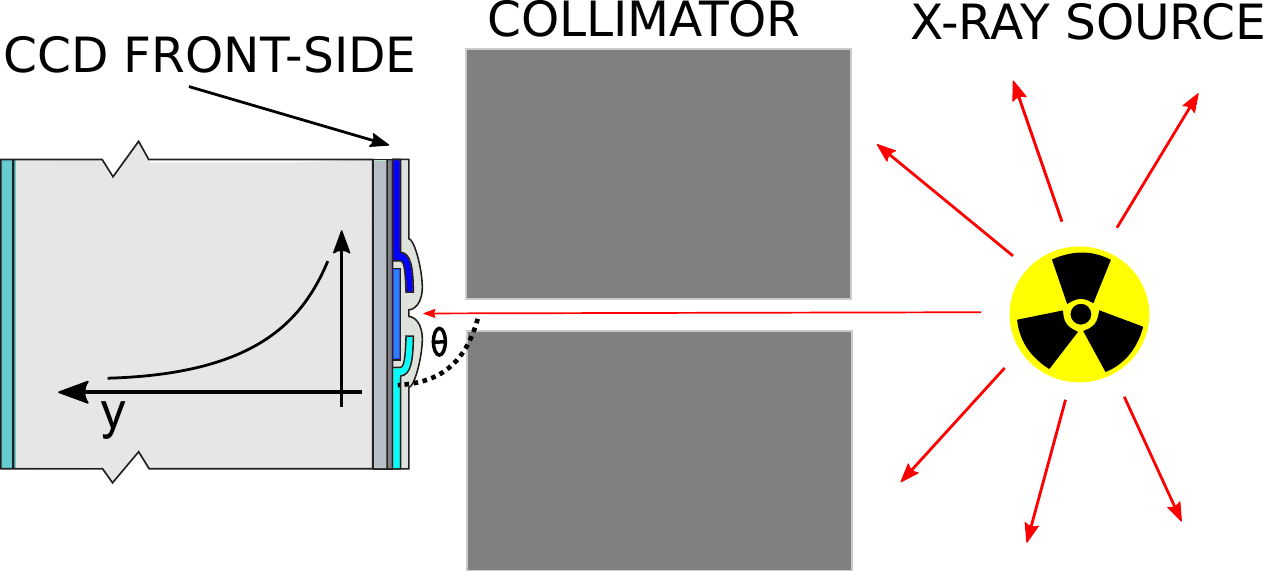}
	\caption{Schema of the experiment. A fraction of the X-rays emitted by the source are collimated arriving perpendicularly to the front-side CCD $\theta=90^\circ$. In the CCD bulk there is an schema of the attenuation distribution $g_Y(y)=\frac{1}{\gamma}e^{-\frac{y}{\gamma}}$ of the X-rays.}
	\label{fig:experimentoColimador}
\end{figure}

Figure \ref{fig:experimentoColimador} shows an schema of the experiment. In this experiment, the CCD front-side was exposed to collimated X-rays. The collimator was a aluminum sheet 10\,mm thick covering the whole CCD, with a pinhole of 0.5\,mm diameter to collimate the X-rays. The distance between the CCD and the sheet was less than a millimetre. The incident photon reaches the sensor at a straight angle ($\theta = \rm 90\,^\circ$), and therefore its probability of interaction is $g_Y(y)=\frac{1}{\gamma}e^{-\frac{y}{\gamma}}$. 

\begin{figure}[t]
	\centering
	\includegraphics[width=\linewidth]{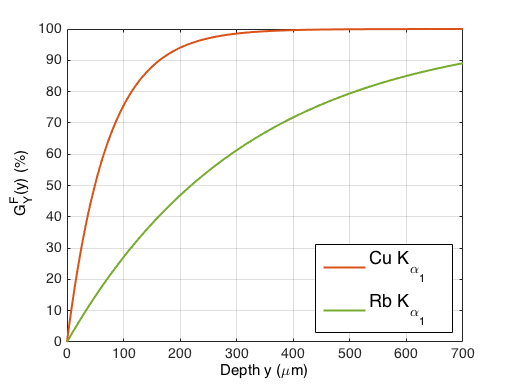}
	\caption{Cumulative distribution $G_Y(y;\theta)$ of the $K_{\alpha_1}$ lines from copper and rubidium fluorescence for an incident angle $\theta = 90^\circ$.} 
	\label{fig:atenuacionXrays}
\end{figure}

Two different photon energies were used in order to test the effect of repulsion for two different number of carriers. These X-rays are the $K_{\alpha_1}$ fluorescence line from copper and rubidium materials. Table \ref{tab:rayosX} shows the energy, attenuation length and average ionization produced in silicon at 140\,kelvin. Using eq. \ref{eq: cumulative distribution from density}, $G_Y(y)$ can be computed for each photon energy which is show in Figure \ref{fig:atenuacionXrays}. 

\begin{table}[]
	\centering
	\caption{Energy and attenuation length of the X-rays used in the experiment \cite{williams2001x}. The ionization was obtained using the factor at 140\,kelvin of 3.77\,${\rm eV/e^-}$ \cite{groom2004temperature}. The last column is the total number of X-rays detected with the CCD.}
	\label{tab:rayosX}
	\begin{tabular}{|c|c|c|c|c|}
		\hline
		X-ray & \begin{tabular}[c]{@{}c@{}}Energy\\ (eV)\end{tabular} & \begin{tabular}[c]{@{}c@{}}Ionization $({\rm e^-})$\end{tabular} & \begin{tabular}[c]{@{}c@{}}$\gamma$\\ $({\rm \mu m})$\end{tabular} & \begin{tabular}[c]{@{}c@{}} Number of \\ events \end{tabular} \\ \hline
		Cu $K_{\alpha_1}$ & 8047,78    & 2152 & 70,8  & 1540 \\ \hline
		Rb $K_{\alpha_1}$ & 13395,30   & 3582 & 316,6 & 2555 \\ \hline
	\end{tabular}
\end{table}

\subsection{CCD sensor under use}
\label{sec:ccd sensor experimental}
The CCD used in this work was cooled at 140\,kelvin to minimize dark current \cite{holland2003fully}. A pixel readout time of 50\,$\mu s$ was selected to achieve a noise level of $\sim$2\,$\rm e^-$. The thickness of these detector is 250 $\mu$m, and it was fully depleted using a substrate voltage of 40\,V. The expected electric field model is obtained from the designer group publication in \cite{holland2003fully}: $N_D\approx6\times10^{11} cm^{-3}$, $y_J\approx0.7\,\mu m$ and $E_J\approx2950\,V/cm$ (for a substrate voltage of 40\,V), we get an electric field constants (from eq. \ref{eq:a_1} and \ref{eq:a_2}) $a_1=92800\,V/cm^2$ and $a_2=2940\,V/cm$. 

\subsection{Spread estimation error for X-ray events}

\begin{figure}[t]
	\centering
	\includegraphics[width=\linewidth]{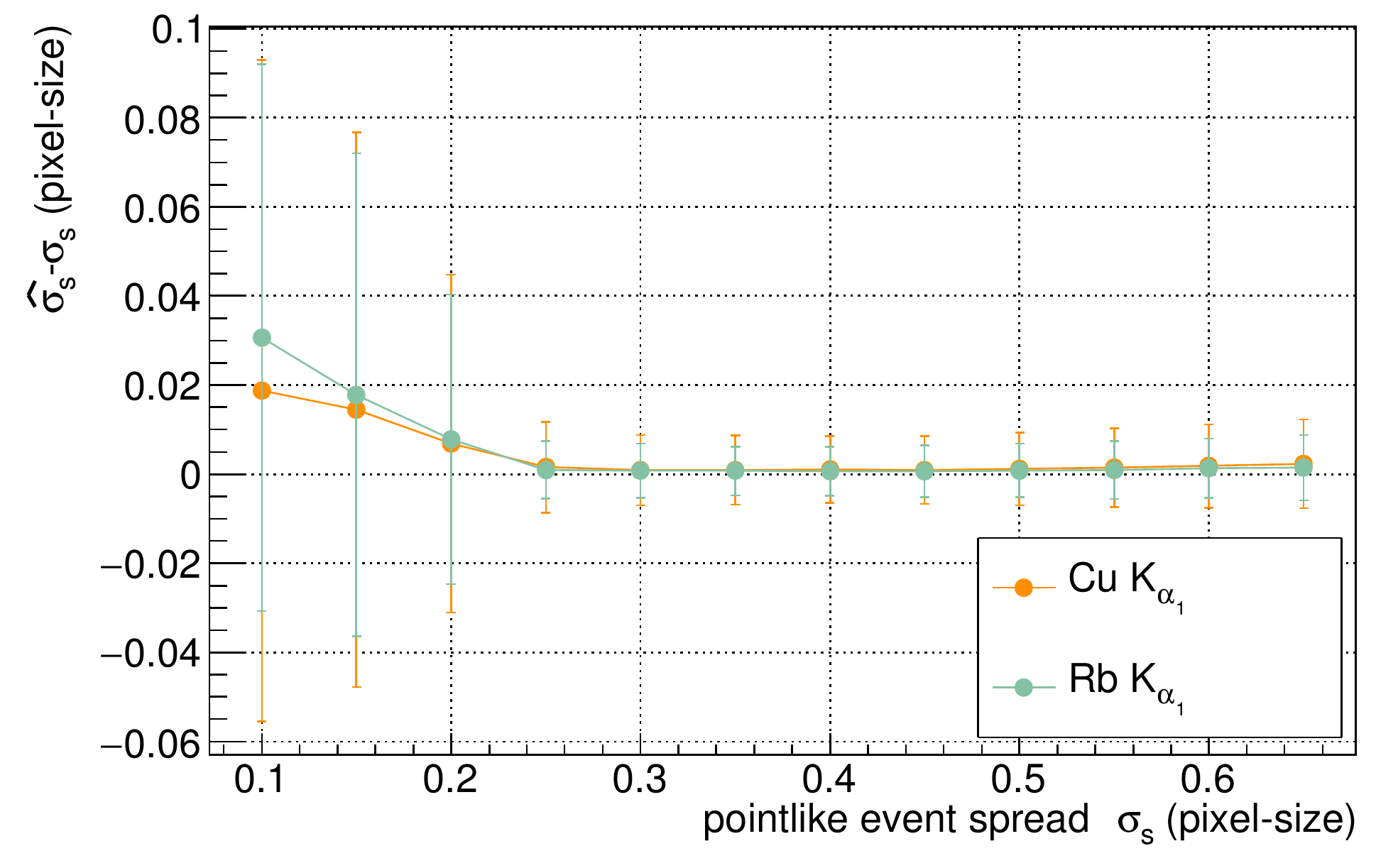}
	\caption{Estimation error of the pointlike event spread $\sigma_s$. For details see text.} 
	\label{fig:errorSpread}
\end{figure}

To evaluate the estimation error of the spread-measurement algorithm in eq. \ref{eq:likelihood}. Events with the charge packets with the number of holes expected for both X-rays energies 2152\,$\rm e^-$ (Cu-line) and 3582\,$\rm e^-$(Rb-line), were simulated using the algorithm in section \ref{sec:simulation} and \ref{sec:ccd sensor experimental}. The lateral original coordinates $\rm (\mu_x,\mu_z)$ of the simulated event was randomly generated with a uniform distribution in the array. Figure \ref{fig:errorSpread} shows the absolute error in the spread reconstruction of the carriers $\hat{\sigma}_s$ in unit of pixels. In events with spread $\rm \sigma_s<0.25$\,pixel-size (as the example in Fig. \ref{fig:ejEventTop}) all carriers are trapped by one or two pixels loosing much of the spatial information of the carriers. For events with $\rm \sigma_s>0.25$\,pixel-size (as the example in Fig. \ref{fig:ejEventBot}), the estimator works in similar error for both X-ray sizes and with a spread estimation resolution below 0.01\,pixel.

\subsection{Depth-spread measurement} \label{sec:xrays}

\begin{figure}[t]
	\centering
	\begin{subfigure}[t]{\linewidth}	
		\centering	
		\includegraphics[width=0.8\linewidth]{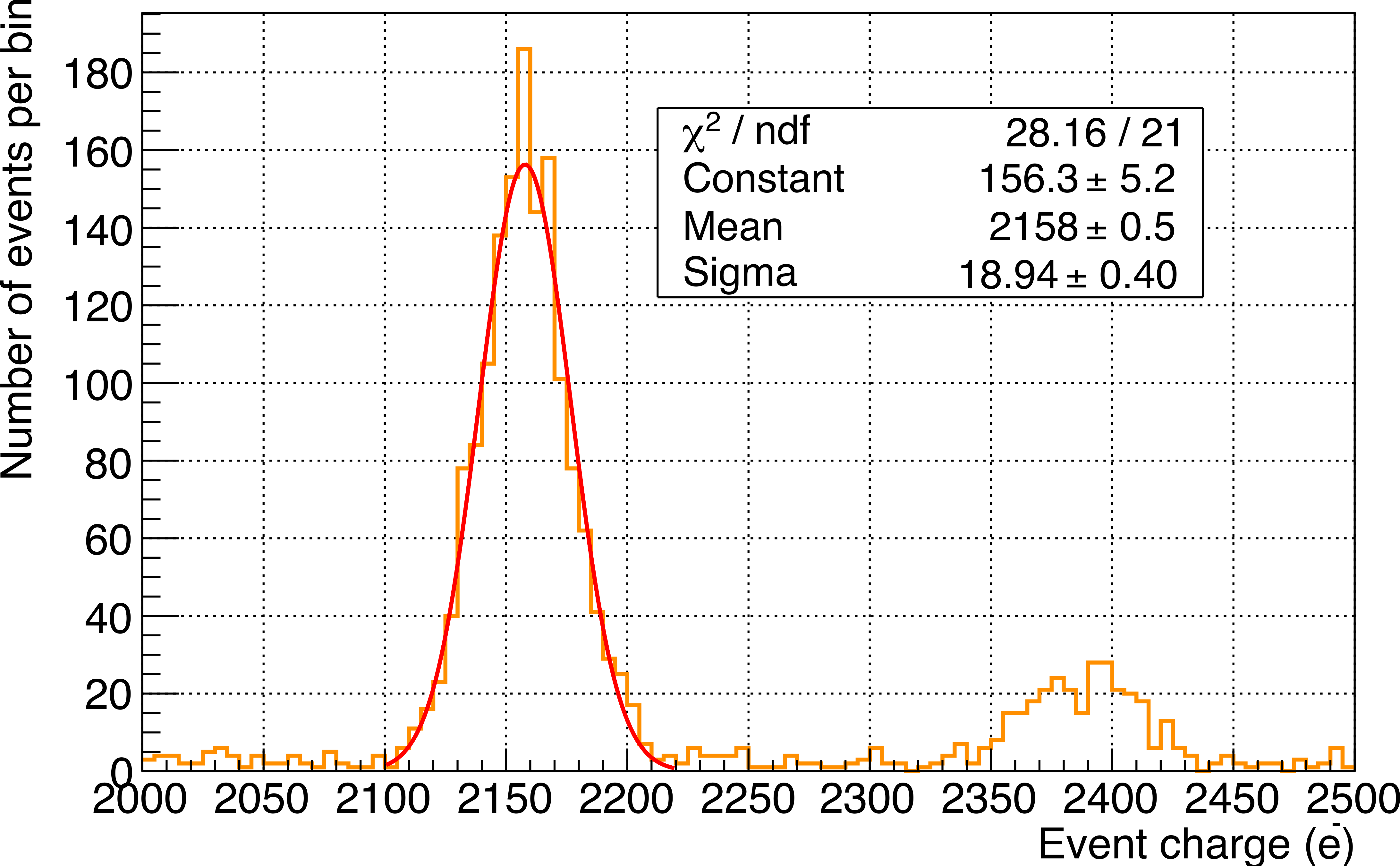}
		\subcaption[]{The peak at 2158\,$\rm e^-$ is produced by the $K_\alpha$ Cu fluorescence X-ray of the source, and the peak at $\sim$2380\,$\rm e^-$ is the $K_\beta$ X-ray.}	
		\label{fig:cuPeak}
	\end{subfigure}\\
	\begin{subfigure}[t]{\linewidth}
		\centering			
		\includegraphics[width=0.8\linewidth]{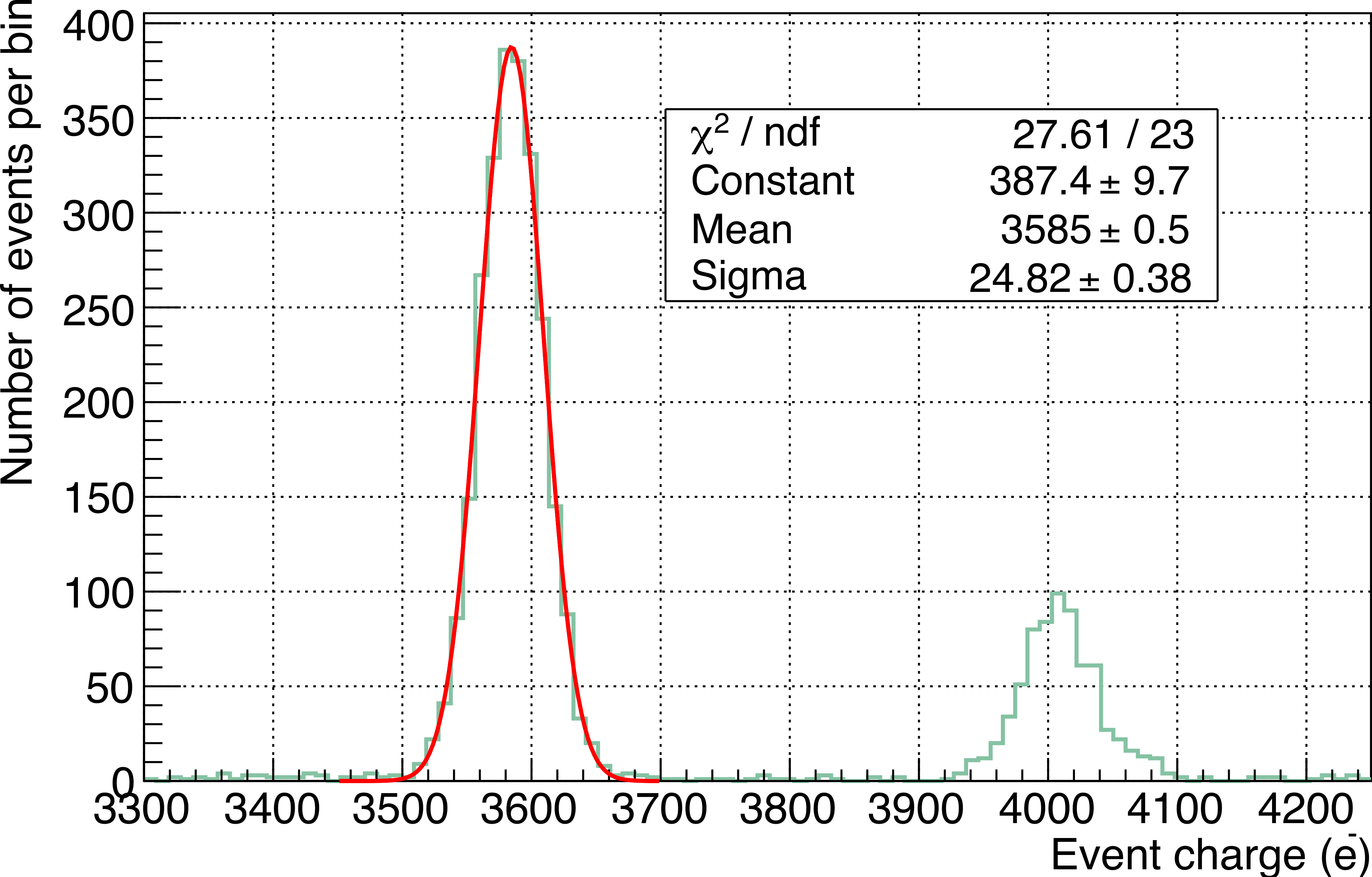}
		\subcaption[]{The peak at 3585\,$\rm e^-$ is produced by the $K_\alpha$ Rb fluorescence X-ray of the source. The peak at $\sim$4000\,$\rm e^-$ is the $K_\alpha$ X-ray from the Itrium of a AlN layer of the CCD package \cite{CONNIEengineRun}.}
		\label{fig:rbPeak}
	\end{subfigure}
	\caption{Charge spectrum of the events recorded with the CCD, at the range of the Cu $K_\alpha$ X-rays peak in (\ref{fig:cuPeak}), and at the range of the Rb $K_\alpha$ X-rays peak in (\ref{fig:rbPeak}). In red is the fit of a Gauss function with the peak data, and the resulting fit parameters are in the box.} 
	\label{fig:espectros}
\end{figure}

\begin{figure}[h]	
	\centering
	\includegraphics[width=\linewidth]{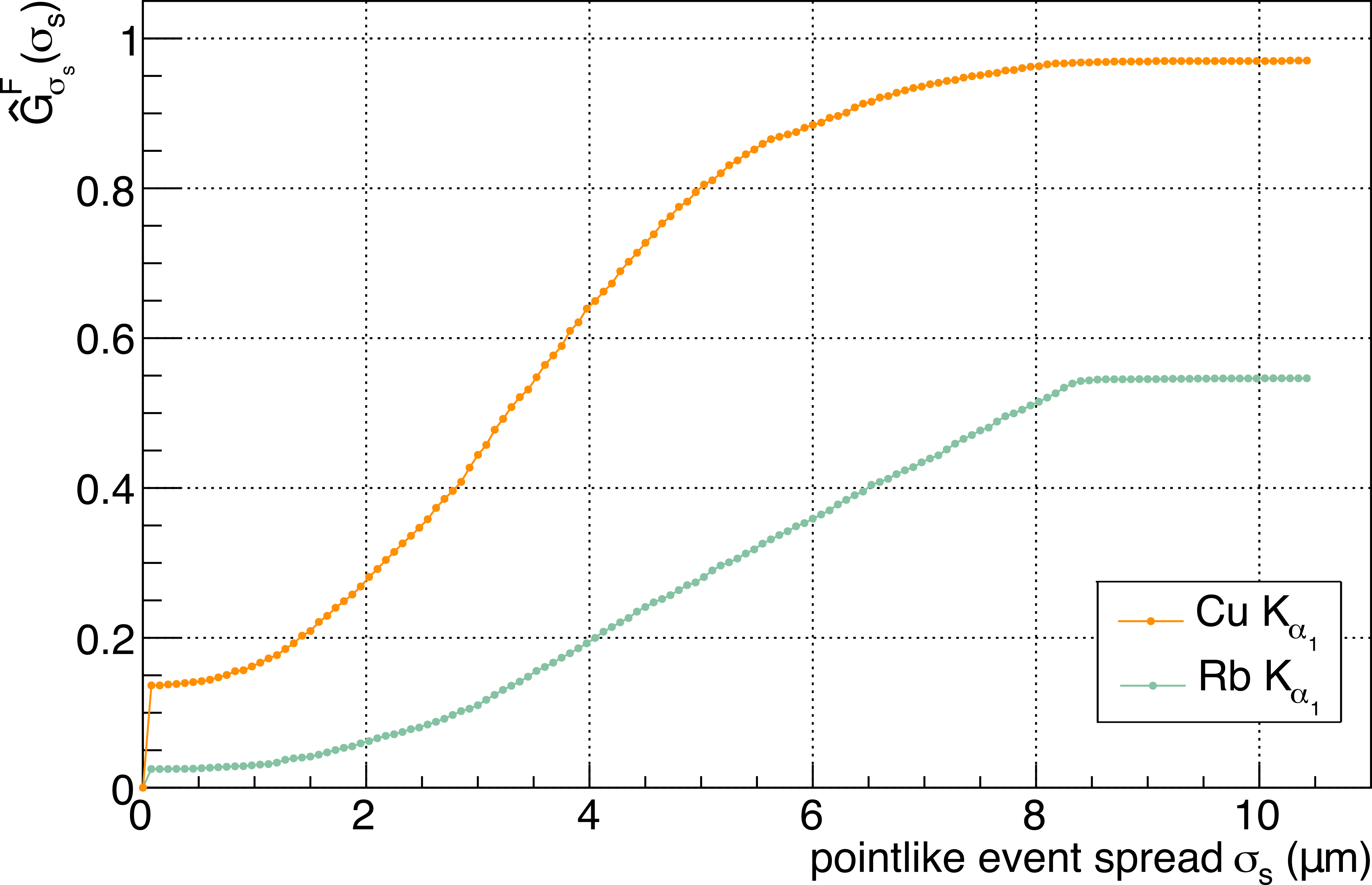}
	\caption{Measured $\hat{G}^F_{\sigma_s}(\sigma_s)$ distribution of each X-ray. As is indicated in equation \ref{eq:cumulative}, they were normalized by $G^F_Y(y_w)$, that is the expected absorbed number of X-rays in the 250\,${\rm \mu m}$ of the CCD silicon.}
	\label{fig:acumuladasSigmaXrays}
\end{figure}

The CCD exposure time was set to avoid pile-up of events in the output image. In Figure \ref{fig:espectros} there are the energy spectrum of the extracted events in the energy range of each X-ray peak. Events three times the peak-sigma around its mean were used for the analysis, therefore more than 99\% of the X-rays are used to apply the method. The charge spread ($\sigma_s$) using the likelihood method in section \ref{sec:mldlh} was estimated for each event. Figure \ref{fig:acumuladasSigmaXrays} is the measured cumulative distribution  $\hat{G}^F_{\sigma_s}(\sigma_s)$ for each photon using eq. \ref{eq:cumulative}.  


\begin{figure}
	\centering
	\includegraphics[width=\linewidth]{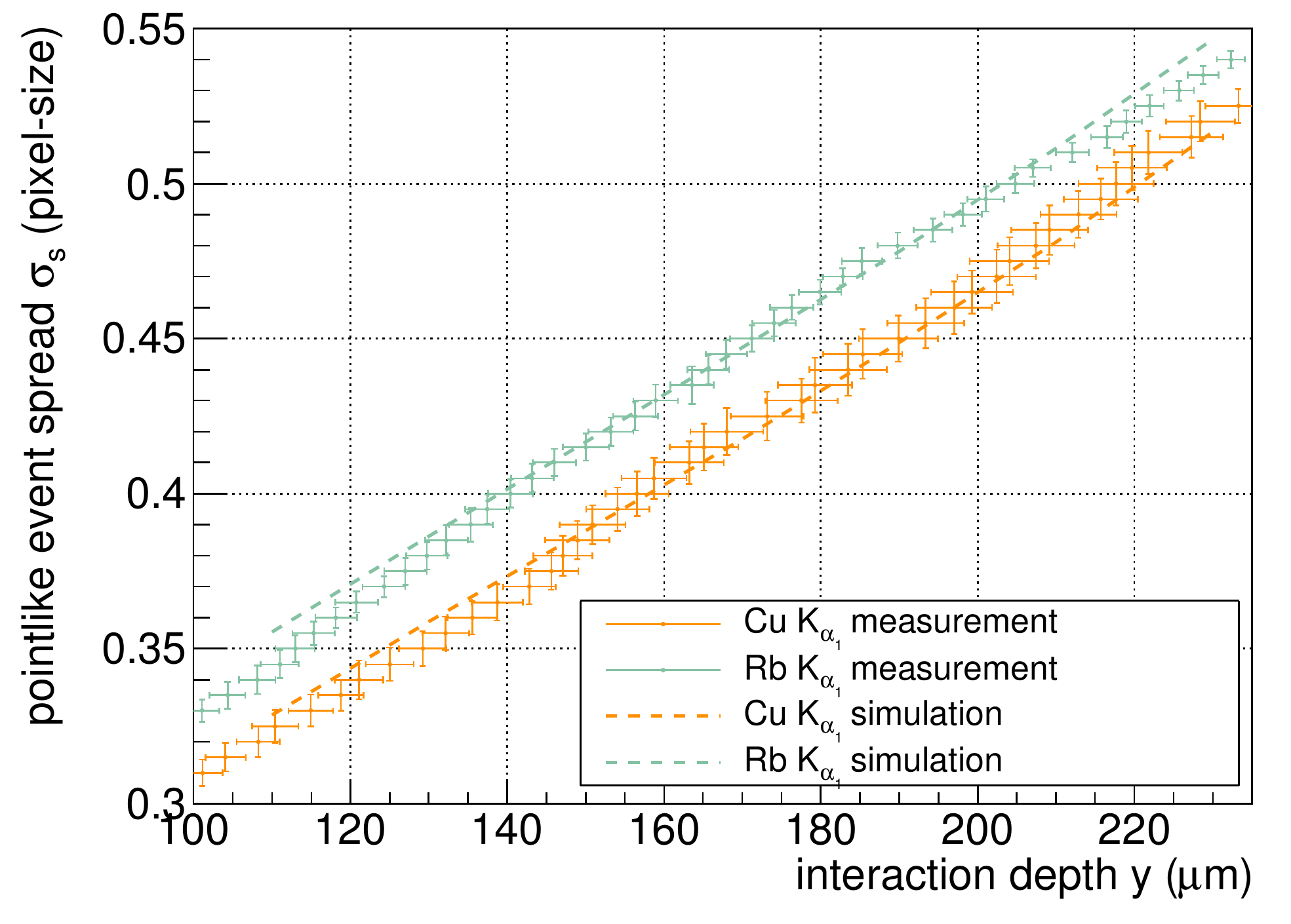}
	\caption{Measured spread-depth function, $\sigma_s(y)$, for charge packets produced by X-rays. In dashed lines are the simulation results.}
	\label{fig:curvas medidas}
\end{figure}

 The curves in Fig. \ref{fig:atenuacionXrays} and \ref{fig:espectros} can be used with eq. \ref{eq:cumulative diffusion} to calculate the $\sigma_s(y)$ function of the CCD. Figure \ref{fig:curvas medidas} shows the resulting measurement. Both photons give different spread for the charge packet generated at the same depth. This behaviour is much larger than the initial cloud size of the interaction and can be explained by the repulsion among carriers: a larger charge packet produces larger repulsion forces that separate the carriers more. The curves were calculated for $y>100\mu$m, because the spread estimator introduces a higher systematic error.

\subsection{Simulation algorithm calibration} \label{sec:simulacion}

To validate the experimental results and to extrapolate the results to other charge packet sizes, we calibrate our simulation algorithm of the charge transport in the CCD bulk. A constant time step of 0.01 ns (validate in \cite{castoldi20123}) was chosen for the simulation. Due to the variability of the CCD fabrication process the electric field parameters ($a_1$ and $a_2$) from eq. \ref{eq:electric field} could change from sensor to sensor. For these purpose a different values of $a_1$ (from 85000 to 108000\,V/cm$^2$ in steps of 1755\,V/cm$^2$) and $a_2$ (from 2600 to 3000\,V/cm in steps of 50\,V/cm) were used in the simulation and compared to results in Fig. \ref{fig:curvas medidas}. For each parameter set a spread-depth function were obtained and its mean root squared error to the measured curves for both X-rays was calculated. The best fit was obtained for $a_1$ = 95777.8\,V/cm$^2$ and $a_2$=2900\,V/cm. The result can be seen in figure \ref{fig:curvas medidas} in dashed lines. They show a good agreement with the measurements and the electric field model is similar the design specification used in section \ref{sec:ccd sensor experimental}. The statistical errors of the measurements were estimated simulating the same experiment many times with the depth cumulative profile from Fig. \ref{fig:curvas medidas} and the number of X-rays events from table \ref{tab:rayosX}.  

\begin{figure}
	\centering
	\includegraphics[width=\linewidth]{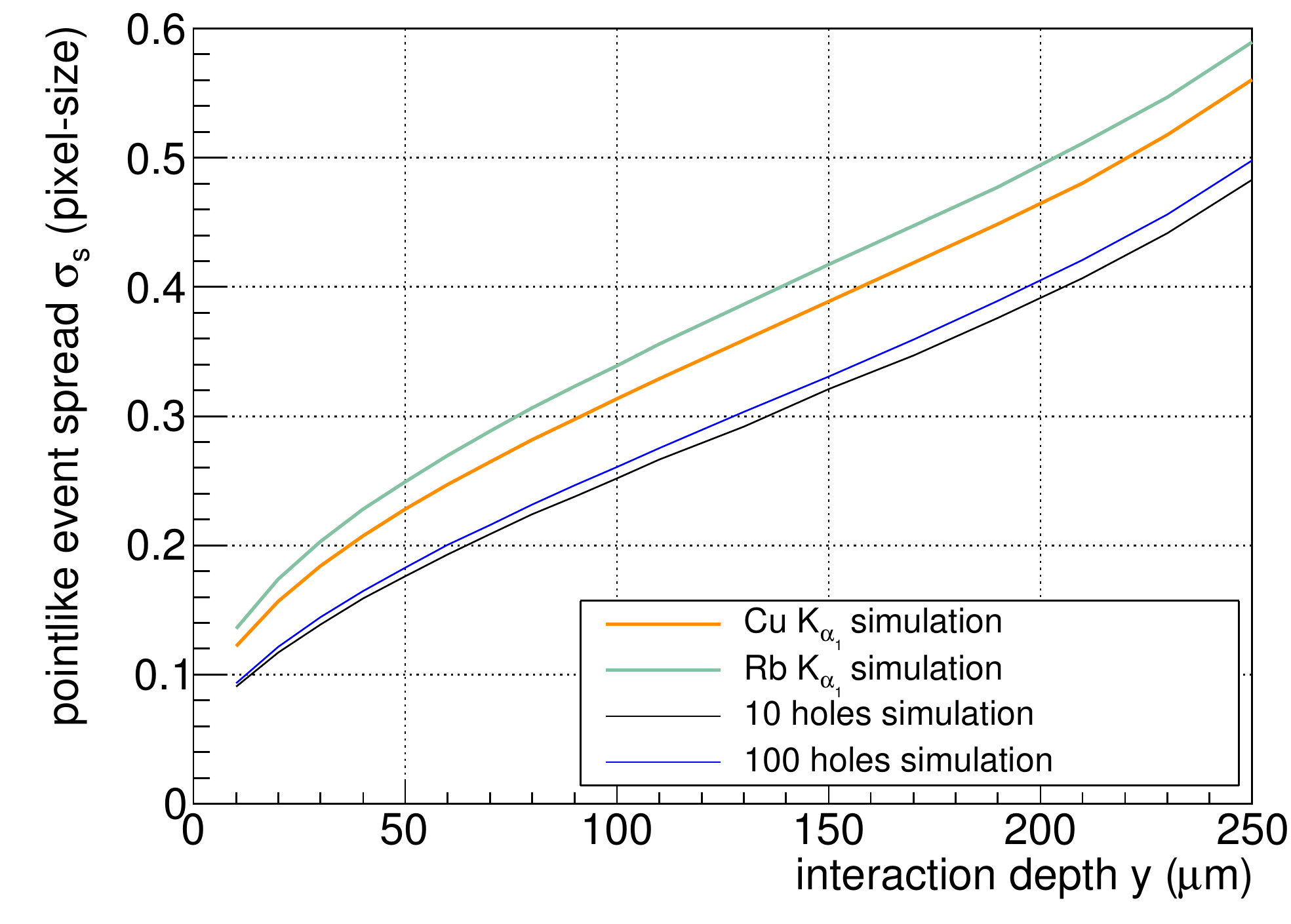}
	\caption{Simulation of the charge transport model for different charge packet values.}
	\label{fig:curvas simuladas}
\end{figure}

As a final step, we use the calibrated simulation to extrapolate the results to very small charge packets which is the region of interest for the ongoing CCD's experiment. The new curves for 10, 100 free carriers are plotted in Fig. \ref{fig:curvas simuladas}. The curves helps to understand the importance of considering the repulsion effect in the calibration and efficiency calculation of pointlike events in the final images.

\section{Conclusions} \label{sec:conclu}

In this work a model of the charge transport of charge packets generated in fully-depleted thick CCDs was presented. Besides the diffusion effect, we also included the Coulomb repulsion of the charge carrier which has not been extensively considered for CCD applications. We presented a new technique to measure the final size of charge packets after its transport through the silicon bulk of the sensor, as a function of the initial ionization point. We applied this new technique in an experiment using collimated X-rays to characterize this relationship. An algorithm to model this transport is presented . The model is fitted to the calibrated data and extrapolated for other charge packet sizes of interest for experiments using the CCDs as particle detectors.


\ifCLASSOPTIONcaptionsoff
  \newpage
\fi


\bibliography{main.bbl}

\end{document}